\documentstyle[12pt,aasms4] {article}
\newcommand{\beq}{\begin{equation}}
\newcommand{\eeq}{\end{equation}}
\newcommand{\beqn}{\begin{eqnarray}}
\newcommand{\eeqn}{\end{eqnarray}}

\newcommand{\non}{\nonumber}

\begin{document}
\title{\bf{A Possible Correlation between the Gaseous Drag Strength and
Resonant Planetesimals in Planetary Systems} }
\author{ Ing-Guey Jiang$^{1}$ and
     Li-Chin Yeh$^{2}$}

\affil{
{$^{1}$ Department of Physics,}\\
{ National Tsing-Hua University, Hsin-Chu, Taiwan} \\
{$^{2}$ Department of Applied Mathematics,}\\
{ National Hsinchu University of Education, Hsin-Chu, Taiwan} 
}

\authoremail{jiang@phys.nthu.edu.tw}

\begin{abstract}

We study the migration and resonant capture of planetesimals 
in a planetary system consisting of a gaseous disc analogous to 
the primordial solar nebula and a Neptune-like planet.
Using a simple treatment of the drag force we find 
that planetesimals are mainly trapped in the 3:2 and 2:1 resonances
and that the resonant populations are correlated with the gaseous 
drag strength 
in a sense that the 3:2 resonant population increases with 
the stronger gaseous drag, but the 2:1 resonant population does not.
Since planetesimals can lead to the formation of larger bodies similar 
to asteroids and Kuiper Belt Objects, the gaseous drag can play 
an important role in the configuration of a planetary system. 
 
\end{abstract}


\newpage
\section{Introduction}

The discovery of more than 160 planets orbiting other stars, 
extra-solar planets, opened a window of astronomy, which could 
eventually 
provide insight to understanding 
the fundamental questions about
our own Solar System.  
Most of these extra-solar planets (exoplanets) were detected 
by the measurements of stellar radial velocities 
of their host star
via the Doppler effect
and several more recent detected exoplanets were discovered
through the transit events. For example, the OGLE project has detected
5 transit exoplanets (Konacki et al. 2003, Bouchy et al. 2005), and 
the TrES team (using 10 cm telescopes)
has 
discovered
one transit planet (Alonso et al. 2004). 
All of the above 6 exoplanets
were 
later 
confirmed by  spectroscopic follow-up.  
Future observations using 
both methods will lead to the 
discoveries of 
additional
systems,
thereby 
improving the statistical significance of the population.

Among the many interesting properties
exhibited by these planetary systems are the large range of
masses, orbital periods
and orbital eccentricities (Jiang et al. 2006).
In particular, the relations between the orbital periods of different planets
have been noted and are 
usually connected with the mean motion resonances.
As an example of extra-solar multiple planetary systems which 
show the resonances,  Ji et al. (2003) confirmed that 
the 55 Cancri planetary system is indeed in the 3:1 mean motion resonance
by both the numerical simulations and secular theory. The 
GJ 876 and HD 82943 planetary systems are probably in 2:1 resonance
as studied by Laughlin \& Chambers (2001), Kinoshita \& Nakai (2001),
Gozdziewski \& Maciejewski (2001) and also Ji et al. (2002).
Moreover, the periods of small bodies in the Solar System 
such as asteroids and Kuiper Belt Objects (KBOs) are also seen 
to have similar connections with resonances. Thus, such types of dynamics may 
also play a role in the configuration of KBOs.

On the other hand, 
it has been shown that discs can affect 
the orbital evolution of test particles
within planetary systems (Jiang \& Yeh 2004a, 2004b; Yeh \& Jiang 2005).
In particular, Jiang \& Yeh (2004c) proposed a possible model of 
resonant capture for proto-KBOs driven by the gaseous drag,
finding that many test particles can be captured into the 3:2 resonance, 
consistent with the observational results (Luu \& Jewitt 2002). 

To ensure that the gaseous drag influences the dynamics of 
planetesimals,
sufficient gas must be present (about $0.01 M_{\odot}$
as suggested in Nagasawa et al. 2000)
when the planetesimals 
are already formed. 
It is likely that molecular gas is present around the 
nearby star epsilon Eridani 
as found by Greaves et al. (1998),
however, only an upper limit of 0.4 Earth masses in 
molecular gas is 
inferred from 
CO observations.  This is to be  
compared to the primordial solar nebula where the 
minimum-mass solar nebula is about 0.026 $M_\odot$.
 
Although there is evidence for a small amount of gas present in 
planetary systems, there may have been much more gas in the past.
The km-sized planetesimals representing  proto-asteroids were  
formed and likely influenced by the gaseous drag.
During the above process,  
the gaseous component is gradually 
depleted from a more massive primordial nebula
to the current limited molecular gas. 

Whether the above scenario is viable would be related to the formation
time-scale of km-sized planetesimals and  the depletion time-scale of the 
gaseous discs. Cuzzi et al. (1993) argued that 10-100
km sized objects can be formed in about $10^6$ years.
Furthermore, observations by Kenyon \& Hartmann (1995), 
Haisch et al. (2001) show that at the age of about $10^6$ years, most low-mass
stars are surrounded by the optically thick discs. However,
by the age of $10^7$ years, no such discs are detected.
It is therefore possible 
that there is a phase in the evolution of the system
where planetesimals are already 
formed while the gaseous disc is not yet depleted.
We investigate this phase to examine the effect of  
gaseous drag for the planetesimal 
dynamics in this paper. 

Since the km-sized planetesimals
will grow into asteroids, KBOs or even planets,
their distribution and orbital evolution
are extremely important for understanding the history of planetary systems. 
Based on a disc model analogous to the primordial solar nebula,
we study the resonant capture of planetesimals under the influence
of a gaseous disc for a given planetary system. In particular, we 
study 
the possibility for  
correlations between the gaseous drag strength and the resonant populations 
and examine 
the possible parameter space for  which the  
drag-induced resonant capture
could explain the resonant populations of a planetary system.

We present our model and assumptions in $\S$ 2.
$\S$ 3 describes the evolution of planetesimals and the
stability tests. 
Finally, we discuss the results and conclude in the last section.

\section{The Model} 



We consider the motion of a test particle influenced by 
the gravitational force from the central star, a planet  
and the proto-stellar disc. This disc, which is mainly composed of gas,  
exerts a drag on 
the planet and the test particles. These test particles are envisioned 
to represent planetesimals such as
proto-asteroids and proto-KBOs.
We assume that the mass of central star
$\mu_1=1 M_\odot$, and 
the planet's mass is taken to be similar to the Neptune's mass, 
i.e. $\mu_2=5\times 10^{-5} M_\odot$. 
We assume the test particle represents a planetesimal with 
a radius about 10 km, with a mass assumed to be about 
$\mu_3=4.3\times 10^{-15} M_\odot$ 
(when the density is similar with the Pluto's).
The coordinates of the central star, 
the planet and the planetesimal are $(0,0)$, $(x,y)$ and  $(\xi,\eta)$,
respectively.

In this paper, we consider the general situation that 
the planet is free to move on any non-circular orbit.

\subsection{The Units}

In this paper, the unit of mass is $M_{\odot}$ and 
the unit of length is 30 AU. Since we set the gravitational constant
$G=1$, the total simulation time, $3.8\times 10^5$, which would correspond to
$10^7$ years.
All simulations start from $t=0$ and terminate
at $t=t_{\rm end}=123200\pi\sim 3.8\times 10^5$.

\subsection{The Equations of Motion}

In this paper, we only consider the coplanar case, 
so all motions are in a two dimensional plane. 
The equations of motion are
\beq\left\{
\begin{array}{ll}
&\frac{d^2 x}{dt^2}=-\mu_1\frac{x}{r^3}-\frac{\alpha}{\mu_2}
\left(\frac{dx}{dt}-v_{x}\right)\rho_N-\frac{dV_N}{dr}\frac{x}{r}, \\
&\frac{d^2 y}{dt^2}=-\mu_1\frac{y}{r^3}-\frac{\alpha}{\mu_2}
\left(\frac{dy}{dt}-v_{y}\right)\rho_N-\frac{dV_N}{dr}\frac{y}{r}, \\
&\frac{d^2\xi}{dt^2}=-\mu_1\frac{\xi}{r_1^3}+\mu_2
\frac{x-\xi}{r_2^3}-\frac{\alpha}{\mu_3}(\frac{d\xi}{dt}-v_{\xi})
\rho_T-\frac{dV_T}{dr_1}\frac{\xi}{r_1}, \\
& \frac{d^2\eta}{dt^2}=-\mu_1\frac{\eta}{r_1^3}+\mu_2
\frac{y-\eta}{r_2^3}-\frac{\alpha}{\mu_3}(\frac{d\eta}{dt}-v_{\eta})\rho_T-
\frac{dV_T}{dr_1}\frac{\eta}{r_1}, 
\end{array}\right.\label{eq:ini1}
\eeq
where
\beq
r^2=x^2+y^2, \label{eq:r}
\eeq 
\beq
r_1^2=\xi^2+\eta^2,\label{eq:r1}
\eeq
\beq 
 r_2^2=(x-\xi)^2+(y-\eta)^2.\label{eq:r2}
\eeq
The equations of motion in terms of x and y
describe the planet's orbit
and the equations of $\xi$ and $\eta$ 
give the orbit of a test particle.
In Eq. (\ref{eq:ini1}), $V_N$ is the disc potential for the planet.
In other words, we define that $V_N=V(r)$ to be the disc 
potential at the
planet's location $(x,y)$ where $r$ is given by Eq. (\ref{eq:r}) and 
$\rho_N=\rho(r)$ is the disc density at the planet's location. 
Similarly, $V_T=V(r_1)$ is the disc potential at the
test particle's location $(\xi,\eta)$ where $r_1$ is given  
in Eq. (\ref{eq:r1}) 
and $\rho_T=\rho(r_1)$ is the disc density at the location of the test 
particle.


The disc is represented by an annulus with inner edge $r_{i}$ and
outer edge $r_{o}$, where  $r_{i}$ and $r_{o}$  are assumed to be 
constants. We choose $r_{i}=1/5$, $r_{o}=5/3$ in this paper.
Since we set the unit of length to be 30 AU, 
$r_{i}$ corresponds to the location 
where the Jupiter was formed approximately and
$r_{o}$ corresponds to the outer edge of the Kuiper Belt. Thus, our choice
of inner and outer edges of the gaseous disc is based on the possible
properties of the proto-solar nebula. 
 
The density profile of the disc 
is taken to be of the form $\rho(r)=c/r^p$, where 
$r$ is the radial coordinate as in Eq. (\ref{eq:r}), $c$ is a constant 
completely determined by the total mass of the disc and $p$ is a natural 
number. In this paper, we set $p=2$ based on the theoretical work by
Lizano \& Shu (1989). This assumption is consistent with one of the models
of the Vega debris disc (Su et al. 2005).
Hence, 
the total mass of the disc is 
\beq
M_{d}=\int^{2\pi}_{0}\int^{r_o}_{r_i}\rho(r')r'dr'd\phi =2\pi c
(\ln r_o-\ln r_i). 
\eeq
In this paper, the disc mass is assumed to be $M_d=0.01$,
which is the same order as the minimum-mass solar nebula 
(0.026 $M_\odot$).  It is also consistent with the observations by 
Beckwith et al. (1990) in which the discs have masses ranging from 0.001 to
1 $M_\odot$.
The disc's gravitational potential and force can be calculated by 
elliptic integrals as in 
Jiang $\&$ Yeh (2004b).

In Eq. (\ref{eq:ini1}), 
the Keplerian velocity of the gaseous material in the 
$\xi$ direction is $v_\xi=-\sqrt{\frac{\mu_1}{r_1}}\sin\theta_T$
and in the $\eta$ direction is $v_\eta=
\sqrt{\frac{\mu_1}{r_1}}\cos\theta_T$,
where $\theta_T=\tan^{-1}(\eta/\xi)$. Similarly, 
$v_x=-\sqrt{\frac{\mu_1}{r}}\sin\theta_N$ and  
$v_y=\sqrt{\frac{\mu_1}{r}}\cos\theta_N$, where $\theta_N=\tan^{-1}(y/x)$.

\subsection{The Drag}


In this system, we include the effect of the drag
force from the gaseous disc. There are various forms 
for the drag force that
could be employed. For example, Murray (1994) considered a general drag force
per unit mass of the form:
\beq
{\bf F}=k{\bf V}g,
\eeq
where $k < 0$, ${\bf V}$ is the particle's velocity in the inertial frame,
and $g$ is a scalar function of its position and velocity.

On the other hand,
the aerodynamic friction force per unit mass is given by 
(Fitzpatrick 1970, Fridman \& Gorkavyi 1999)
\beq
{\bf F}= \frac{1}{2m} C_{D} A \rho V_r^2,
\eeq
where $C_D$ is the drag coefficient, $m$ and $A$ are the particle's 
mass and cross-section, $\rho$ is the gaseous density, and
$V_r$ is the magnitude of particle's velocity 
relative to the gas. 

The Epstein drag force per unit mass has also been adopted 
(see Youdin $\&$ Shu 2002, Youdin $\&$
Chiang 2004) and is given by 
\beq
{\bf F}= \frac{4\pi}{3m} \rho_g c_g v a^2,
\eeq
where $\rho_g$ is the mass density of gas and $c_g$ is the gas sound speed.
The particle's mass and size are $m$ and $a$ respectively.
The relative velocity between a particle and gas is $v$. 

Our disc is mainly composed of gas and 
we wish to approximate 
the disc's gaseous drag 
acting on
the planetesimals and planets. 
Motivated by the above expressions for the drag,
we employ a unified formula which we use for both planetesimals
and the planet. 
Thus, we assume that the drag force per unit mass is:
\beq
{\bf F}=-\frac{\alpha \rho}{m} {\bf V},
\eeq
where $\alpha$ is assumed to be a constant, ${\bf V}$ is 
the particle's velocity relative to the Keplerian motion of the gaseous disc,
and $\rho$ is the gas density.

Although our drag differs from the Epstein drag, we can see that  
$\alpha$ plays the role of $4\pi c_{g} a^2/3$ of Epstein drag. 
To obtain an estimate of the value of $\alpha$, we (i) use Eq.(2) of 
Youdin \& Chiang (2004) to calculate $c_g$, (ii) use the radius of 
our planetesimal particle, 10 km, to be the value of $a$, and find that 
$4\pi c_{g} a^2/3$ is about $8 \times 10^{-16}$. We, thus, define 
$\alpha_E \equiv 8 \times 10^{-16}$.
Due to the simple treatment of 
the disc drag and the uncertainty in $\alpha$, we  
adopt 3 different values
of $\alpha$: $\alpha_E$, $5\alpha_E$, and $25\alpha_E$.   

   
For the planet, the drag force in $x$ direction is 
$-(\alpha/\mu_2)(dx/dt-v_x)\rho_N$ and thus 
this term appears in the first equation of Eq. (\ref{eq:ini1}).
Similarly, there is a term 
$-(\alpha/\mu_2)(dy/dt-v_y)\rho_N$ appears in the second equation of 
Eq. (\ref{eq:ini1}).
 
For the planetesimals, the drag force in the $\xi$ and $\eta$ directions
are 
$-(\alpha/\mu_3)(d\xi/dt-v_\xi)\rho_T$ and 
$-(\alpha/\mu_3)(d\eta/dt-v_\eta)\rho_T$
in the third and fourth equation of Eq. (\ref{eq:ini1}), respectively.

\subsection{The Simulations}

A number of planetesimals (300) are randomly distributed in a belt region 
$1.1 \le r \le 1.8$ with a uniform number density, where $r$ is the
radial coordinate. Assuming that one planetesimal is located at 
$(\xi, \eta)$, then we set its initial velocity $(v_{o\xi}, v_{o\eta})$ 
to be 
$v_{o\xi}=-(\sqrt{1/r_1})\sin\theta_T$
and $v_{o\eta}=(\sqrt{1/r_1})\cos\theta_T$, 
where $\theta_T=\tan^{-1}(\eta/\xi)$.
Thus, the planetesimals are initially in circular motion.

To study the outcome of different drag strengths, we adopt 3 different
values of the drag coefficient in our simulations. 
That is, $\alpha= \alpha_E$ in model A,
$\alpha= 5\alpha_E$ in model B and $\alpha=25\alpha_E$
in model C. 

The planet is always located at $(1,0)$ initially,
and was set in
circular motion initially for models A, B, and C.
However, in order to test the influence of the planet's orbital eccentricity,
we also consider a model (D) in which the planet moves on an initial 
eccentric orbit 
of $e=0.3$ with drag coefficient chosen to be 
$\alpha=5\alpha_E$. 

To study the resonance captures, 
the orbital semi-major axis and 
eccentricity for all the planetesimals and planet are calculated. 
To make it clear, 
$x_{\ast}$, $y_{\ast}$ 
represent the coordinates of any particle.  
From Murray \& Dermott (1999), we have
\beqn
&& r_{\ast}^2=x_{\ast}^2+y_{\ast}^2, \non  \\
&& v^2=\left(\frac{d x_{\ast}}{dt}\right)^2+\left(\frac{d y_{\ast}}{dt}
\right)^2.\non 
\eeqn
 
Let $h^2=(x_{\ast}\frac{d y_{\ast}}{dt}-y_{\ast}\frac{d x_{\ast}}{dt})^2$, 
then semi-major axis $a$ and eccentricity $e$ are defined by
\beqn
& &a=\left(\frac{2}{r_{\ast}}-\frac{v^2}{\mu_1}\right)^{-1}, \\
& &e=\sqrt{1-\frac{h^2}{\mu_1 a}}.
\eeqn 
Based on  
the formula in Murray \& Dermott (1999) and Fitzpatrick (1970), the 2:1
resonant argument $\phi_{2:1}$ and 3:2 resonance argument $\phi_{3:2}$ are
also calculated as 
\beqn
& &\phi_{3:2}= 3\lambda_{t}-2\lambda_{N}-\omega_{t}, \\
& &\phi_{2:1}= 2\lambda_{t}-\lambda_{N}-\omega_{t},
\eeqn
where $\lambda_{t}$ is the mean longitude of a planetesimal's orbit,
$\lambda_{N}$ is the mean longitude of the planet's orbit and
$\omega_{t}$ is the longitude of pericentre of a planetesimal's orbit.

The number of particles in a particular resonance at time $t_i$ 
is defined to be 
the total number of particles with 
the difference between the maximum and minimum resonance arguments 
less than $180^o$
during $t_{i-1}< t < t_i$. We set  
$t_0=0, t_i = t_{i-1} + 800\pi$, where $i=1,2,...,154$.

\section{Numerical Results}

\subsection{Evolution of Planetesimals}

The evolution of planetesimals on the $x-y$ plane
in the simulation of model A is illustrated in Fig. 1. 
The panel labelled 0 in Fig. 1 shows the initial positions
of all 300 planetesimals.
Based on this representation,
it is difficult to discern the change in the distribution 
until the 9th panel.
However, specifically it can be seen in
the histograms of particle number versus the radial distance
reveal the variation more clearly 
(see Fig. 2). In the 2nd panel of Fig. 2, a gap starts to develop 
and this gap 
becomes deeper and wider in the following panels. 
Finally, the gap encompasses the
range from 1.5 to 1.7 after the 8th panel. 
The gap can also been seen in panels 10 and 11 of Fig. 1
and separates the planetesimal
distribution into two rings. 
The outer ring (from $r=1.7$ to $r=1.8$) is thinner than the inner one
(from $r=1.1$ to $r=1.5$) as one can see from Fig. 1 and 2. 

In Fig. 3, we plot the final planetesimal distribution of model A
in the $a-e$ plane
on the left panel and also the color contour of it on the right panel.
There is a population with nearly circular orbits from $a=1.7$ to $a=1.8$.
They are the main population of the outer ring of Fig. 1 and 2 as one
can estimate the total number from the color bar.
From both panels and also the color bar, we find that some planetesimals
are associated with 2:1, 3:2, 7:5, 4:3, 5:4, 6:5 resonances  and
that there are more  planetesimals
in 3:2 (at $a=1.33$) than those in 2:1 resonance (at $a=1.6$).
However, in order to confirm whether a given planetesimal is captured into
a particular resonance, the particular resonance argument has to be calculated 
during the whole simulation. Because the influence of the 
first order resonance, 2:1 and 3:2, 
are much stronger than the others, we 
calculate the 2:1 and 3:2 
resonance arguments, i.e. $\phi_{2:1}$ and $\phi_{3:2}$.
Fig. 3 also shows that 
there is a non-resonant population (from $a=1.4$ to $a=1.5$) with nearly
circular orbits.
This population, 
together with all the resonant population between 1.1 and 1.5,
become the main population of the inner wider ring of Fig. 1 and 2.
The orbital eccentricities of the 2:1 resonant population are larger
and thus these planetesimals
have wider range in semi-major axis. Although 
this wider range might cover both rings and also 
the gap of Fig. 1 and 2, the number of
this population is small so that they do not change the distribution.


The evolution of planetesimals' distribution for model B
is plotted in Fig. 4. The evolution 
proceeds more rapidly than 
in model A. This could be due to a larger drag force for this model.
For example, in the 2nd panel, there is already a gap in the belt.
The planetesimals continue to migrate inward and the gap becomes  
wider as shown in panels 3 and 4. In the 5th panel, the planetesimals are 
seen to distribute into
two parts, i.e. an inner ellipse and an outer ring.
The inner ellipse 
precesses slowly
starting from panel 6 until the end of the simulation.
This behaviour is confirmed in the 
histograms in Fig. 5 as the gap begins to form 
in panel 2. The inward migration is also evident as
particle number increases between $r=1$ and $r=1.5$. 

It can be seen in Fig. 6 that 
some planetesimals associated with the 2:1 and 3:2 resonances
cluster around $a=1.6$ and $a=1.33$. 
All the planetesimals associated with the 3:2 resonance have almost exactly 
the same semi-major axes and, moreover, there is no other non-resonant 
planetesimals around 
$a=1.33$, explaining why the inner ellipse is so thin in Fig. 4.


In model C, the largest drag force is considered. 
The planetesimals'  inward migrations 
are so fast that there is already a gap in the 1st panel of Fig. 7.
A clear inner ellipse is formed in panel 2 and it precesses slightly at the 
following panels. With that small precession, the whole distribution seems
to be a steady state. The histograms of Fig. 8 
also confirm the existence of the gap. 
We point out 
that there are no planetesimals in 
the 2:1 resonance for this model (see Fig. 9). 

Although the planet's orbital
eccentricity is taken to be zero initially, its eccentricity 
is about 0.02 during the simulations for models A, B, and C.  
In model D, the strength 
of the drag force is assumed to be moderate, 
i.e. the same as the one in model B, 
but the initial
orbital eccentricity of the planet is 0.3. 
Due to the drag, the inward migrations are still significant and 
the gap immediately appears in the 1st and 2nd panel of Fig. 10. 
However, 
planetesimals continue to migrate inward
from panel 3 to 11 until they arrive at the inner edge of the disc. 
Finally, there is a small ring 
at the disc's inner edge and the planetesimals at the outer regions are 
distributed
randomly. Indeed, the histogram in Fig. 11 shows that most planetesimals
are at the inner edge. 


We summarize the results on the number of planetesimals 
captured into 3:2 and 2:1 
resonances in Fig. 12.
For model A,
the solid curves show that there are about 50 planetesimals captured 
into 3:2 resonance and about 40 into 2:1 resonance. 
It is also shown in panel b that the number of the 2:1 resonant planetesimals
approaches the 
maximum earlier than the one of 3:2. 
On the other hand, for model B,
the dotted curve shows that there are about 140  planetesimals
captured into the 3:2 resonance. It is much more than the
one for model A. However, The panel b shows that
the number of planetesimals captured into the 2:1 resonance in model B 
is the same as the one in model A.  
Moreover,
since there are no planetesimals in 
the 2:1 resonance for model C,
there is no dashed curve in panel b.
For model D,
the long dashed curves  
show that only a few planetesimals
are captured into the 3:2 and 2:1 resonances. 
Therefore, 
the planet's eccentric orbit significantly reduces the 
probabilities of resonant captures.

\subsection{The Stability Tests}

To determine the stability of the results by perturbations in the initial
orbital eccentricity, disc mass, and location of the inner and outer 
disc edge, we have carried out a simulation consisting of 30 
planetesimals with parameters identical to those adopted in model B.
The limited number of planetesimals used here allows one to carry out 
many tests quickly. 
In this standard simulation, all 30 planetesimals 
migrate inward and are captured into the 3:2 resonance. 

To check the stability of the orbital evolution to  the 
initial orbital eccentricities of 
planetesimals, a simulation was performed in which all planetesimals have
initial 
orbital eccentricities $e=0.01$ with all other settings remaining 
the same as the standard one.
These 30 planetesimals migrate inward of which 
27 are captured into the 
3:2 resonance. Because the difference with the standard case 
is only 3 planetesimals, which is one order of magnitude
less than the total number, 30, we conclude that the system is stable in
terms of the initial orbital eccentricities of planetesimals.
For the results  in which 300 planetesimals are used,
we are confident that the result would be similar if the planetesimals'
initial orbital eccentricities were changed slightly.

To test the stability in terms of the disc mass, i.e. the value of $M_d$,
we ran two simulations, one with $M_d=0.009$ and 
another one with $M_d=0.011$
while keep all other settings the same as the
standard one. Note that $M_d=0.01$ in the standard case.
For both simulations, 
all the 30 planetesimals migrate inward and are captured into
the 3:2 resonance. Thus, the system is stable 
to perturbations 
in the disc mass.

To check the stability in terms of the location of the disc's 
edges, i.e. the 
value of $r_i$ and $r_o$, 4 simulations were carried out
with $r_i=1/5 - 0.01$, $r_i=1/5 + 0.01$,
$r_o=5/3 - 0.01$, $r_o=5/3 + 0.01$, 
 while 
all other settings remaining the same as the
standard case. 
All the 30 planetesimals migrate inward and are captured into
the 3:2 resonance for all 4 simulations.
Hence, the system is stable in terms of 
perturbations to the disc inner and outer edges.

\section{Discussions and Conclusions}

In this paper, 
we have investigated the effect of different strengths of the gaseous drag  
on the resonant capture into the 3:2 and 2:1 resonances. 
For a small drag force as in model A, 
there are about 17 percent of planetesimals
captured into the 3:2 resonance and about 
13 percent into the 2:1 resonance. For a moderate drag force as 
in model B, the fraction of planetesimals captured into 2:1 is still about 
13 percent
but the fraction for the 3:2 resonance increases to 47 percent.
When a stronger drag force is used as in model C, the 
fraction of planetesimals
captured into the 3:2 resonance greatly increases up to about 60 percent. 
In contrast,
the number for the 2:1 resonance becomes zero. 
Therefore, the numerical results of the resonant capture process reveal that 
it is very 
sensitive to the strength of the gaseous drag. Since the planetesimals
are captured into resonances during their inward migrations,
the stronger drag increases the speed of inward migration,
so, equivalently, the resonant population in capture processes is correlated
with the speed of inward migration. 

For a model in which the planet has an initially eccentric orbit, less than
3 percent of the planetesimals are trapped into the 3:2 and 2:1 resonances. 
Hence, the assumption of a large finite eccentricity nearly destroys the 
possibility of resonant captures.



To understand the difference between the 3:2 and 2:1 resonant captures,
there are mainly two possibilities:

(1) the details of capture processes for the 3:2 and 2:1 resonances 
    are fundamentally different, strongly depend on the migration
    speed of planetesimals. 
 
(2) under our assumptions, there are more planetesimals
    passing the 3:2 resonant region during the simulations, so that
    more planetesimals are captured into the 3:2 resonance, even though
    the capture probabilities for 3:2 and 2:1 resonances are similar. 
    
In order to understand whether the 2nd explanation
 is correct, we estimate the
capture probability for both the 3:2 and 2:1 resonances. 
Let us assume that the potential candidates to be captured into 
the 2:1 resonance are  
those planetesimals initially located between the outer edge ($r = 1.8$) and 
the 2:1 resonant region (about $r = 1.6$). 
Since there are $300$ planetesimals uniformly distributed in 
the region $1.1 \le r \le 1.8$, the potential number of planetesimals
to be captured into the 2:1 resonance is 
$$300\times \frac{(1.8^2-1.6^2)}{(1.8^2-1.1^2)}\sim 100.$$
For model A, 
the total number of planetesimals 
captured into 2:1 is about 40, and the capture
probability for the 2:1 resonance is about $40\%$.

The potential candidates to be captured into 
the 3:2 resonance, on the other hand,  
are those planetesimals initially located between the 2:1
resonant region (about $r = 1.6$) and 3:2 resonant region 
(about $r = 1.33$) plus 
those planetesimals initially located out 
of $r = 1.6$ but {\it not} captured into the 2:1 resonance, so 
the potential number 
of planetesimals to be captured into the 3:2 resonance is    
$$300\times \frac{(1.6^2-1.33^2)}{(1.8^2-1.1^2)} + 60 \sim 177.$$
Thus, 
the capture probability for the 3:2 resonance is about $50/177=28\%$, 
which is smaller than the one for the 2:1 resonance.
    
For model B, the capture probability for the 2:1 resonance
is still about $40\%$, however the total number of planetesimals captured into
the 3:2 resonance 
increases significantly up to about 140. This could be due to the fact that
some planetesimals initially 
located out of $r=1.6$  but {\it not} captured into the 2:1 resonance
are also captured into the 3:2 resonance in time due
to fast inward migrations. 

Since the potential number of 
planetesimals to be captured into the 3:2 resonance is again $177$,
the capture probability for the 3:2 resonance is $140/177$, 
which
is about $79\%$. Therefore, not only is the number of planetesimals 
passing into
the 3:2 resonant region larger than the one for the 2:1 resonance, but also 
that the capture
probability of the 3:2 resonance is larger than that of the 2:1 resonance.

For the largest 
gaseous drag considered (model C),
the speed of inward migration is
highest and even more planetesimals are in the 3:2 resonance. 
However, in this case, 
the number of particles in the 2:1 resonance vanishes.

From the above analysis for models A, B, and C, 
we confirm that the details of the 3:2 and 2:1
resonant capture are fundamentally different.
As discussed in Peale (1976) 
and Murray \& Dermott (1999),
the resonant relations are determined by the influence
of the planet on the planetesimals.
In particular, during the conjunctions, 
the net tangential force experienced by the 
planetesimal is key because 
it can change the planetesimal's angular momentum. 

In the case when the planet moves on a circular orbit
and the planetesimals are assumed to move on eccentric orbits, 
there is no net tangential force if the conjunctions occur exactly at 
the peri-center or apo-center. 
When the conjunctions occur at any other 
point on the orbit,
the symmetry is destroyed, and thus, there is 
net tangential force. 
For a dense population of 
 planetesimals moving on the same
eccentric orbit, the net  
tangential force would cause  about half of them to gain  angular momentum
and another half to lose angular momentum. 
If all these planetesimals were driven to migrate inwards due 
to the drag, (1) those which 
gain angular momentum would expand their orbits 
a bit and could be captured into the resonance, (2) 
those which lose angular momentum would continue to migrate inward.
Since (a) the planetesimals might move on different orbits
and (b)  
the exact locations of the repeated conjunctions are not known, it 
is difficult to estimate the capture probability.

The symmetry is further destroyed
when the planet moves on the orbit with large eccentricity 0.3.
As a result, it
is more difficult for those conjunctions 
which lead to planetesimals gaining angular momentum
to take place repeatedly.
 Thus,
there are few planetesimals captured into the resonances in model D.

For the Kuiper Belt in the outer Solar System, 
objects are known to occupy the
3:2 and 2:1 resonant regions.  
The conventional mechanism explaining these resonances relies on 
the resonant capture 
by an outwardly migrating Neptune with a migration time-scale 
$\tau= 2 \times 10^{6}$ years (Malhotra 1995).
This migration was assumed to start in the late stages of 
the genesis of the Solar System when the formation of the 
gas giant planet was largely complete, the solar nebula had lost 
its gaseous component, and the evolution was dominated by the 
gravitational interactions.
However, this mechanism is based on 
an assumption of pure radial orbital migrations. 
The generality of this mechanism
is unclear if a more realistic orbit of Neptune were chosen.  
As shown by the numerical simulations in Thommes et al. (1999) and 
the analytic calculations in Yeh \& Jiang (2001), Neptune's orbital
eccentricity shall not be zero during the outward migration.
Because Neptune is currently moving on a circular 
orbit, 
a massive disc is needed to circularize Neptune's orbit 
if the outward migration did happen.

On the other hand, 
our results show that the drag-induced resonant capture can explain the
existence of objects in both 3:2 and 2:1 resonances, 
but the ratio of these two populations
will depend on the gaseous drag strength.
The similarity between the conventional picture and our mechanism 
reflects the fact that both captures are due to the relative motions between 
the planet and the small bodies. The main difference is the causes of the 
relative motions.  


Finally,
our results (model D) also show that  
the resonant capture occurs provided that 
the planet's orbital eccentricity  is not too large. 
This result
could place some constraint on the possible orbital history of the planet.
For example, from this point of view, 
in the conventional picture, Neptune will be able to capture 
the KBOs into the resonances only when 
its eccentricity is reduced to be smaller than 0.3. 
This further constrains the orbital history and the timing of Neptune's
migration  if it did significantly
contribute 
on the resonant capture of 3:2 and 2:1 resonant KBOs.
 
\section*{Acknowledgment}
We acknowledge the anonymous referee's many good suggestions. We especially
thank Prof. R. Taam for his efforts in improving  
the presentation. 
We are grateful to the National Center for High-performance Computing
for computer time and facilities. 

This work is supported in part 
by the National Science Council, Taiwan, under 
Ing-Guey Jiang's Grants: NSC 94-2112-M-008-010 and also 
Li-Chin Yeh's Grants: NSC 94-2115-M-134-002.

\clearpage
\begin{figure}[tbhp]
\epsfysize 6.5 in \epsffile{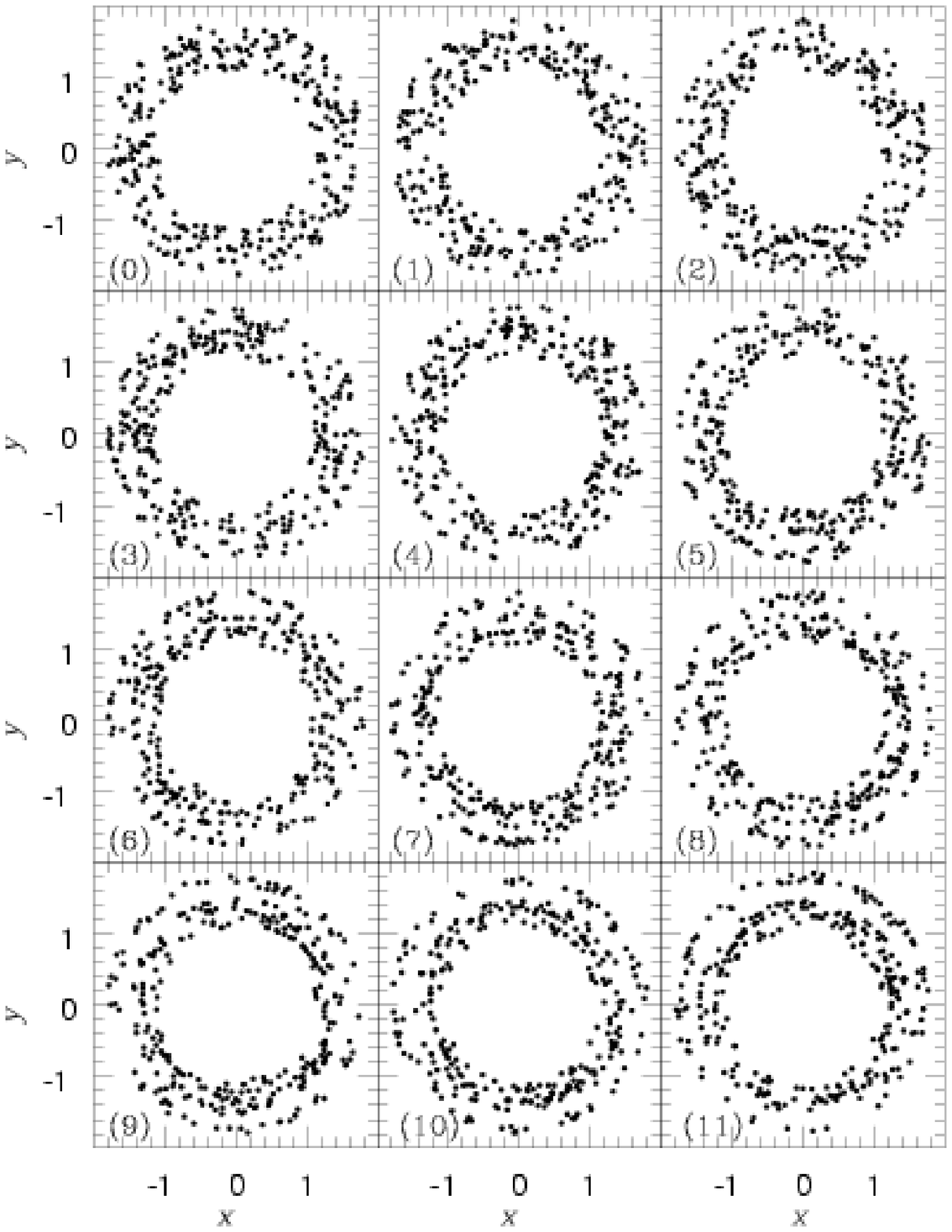}
\caption{The model A: the evolution of particle distributions in the 
$x-y$ plane.
The time between successive panels corresponds to $11200\pi$.}
\end{figure}
\clearpage
\begin{figure}[tbhp]
\epsfysize 6.5 in \epsffile{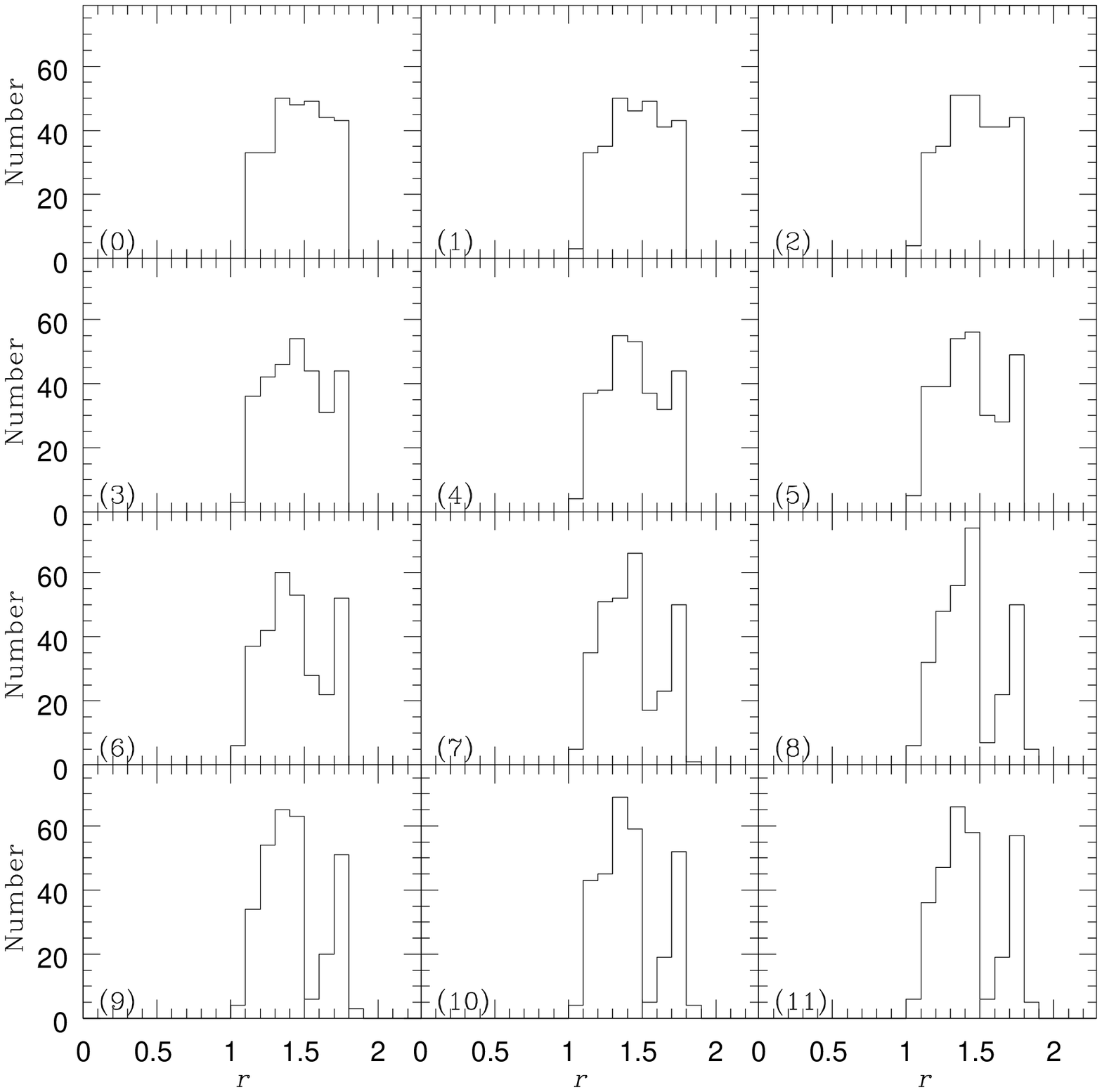}
\caption{The model A: the histograms of particle distributions 
in the radial coordinate.
The time between successive panels corresponds to $11200\pi$.}
\end{figure}
\clearpage
\begin{figure}[tbhp]
\epsfysize 3.0 in \epsffile{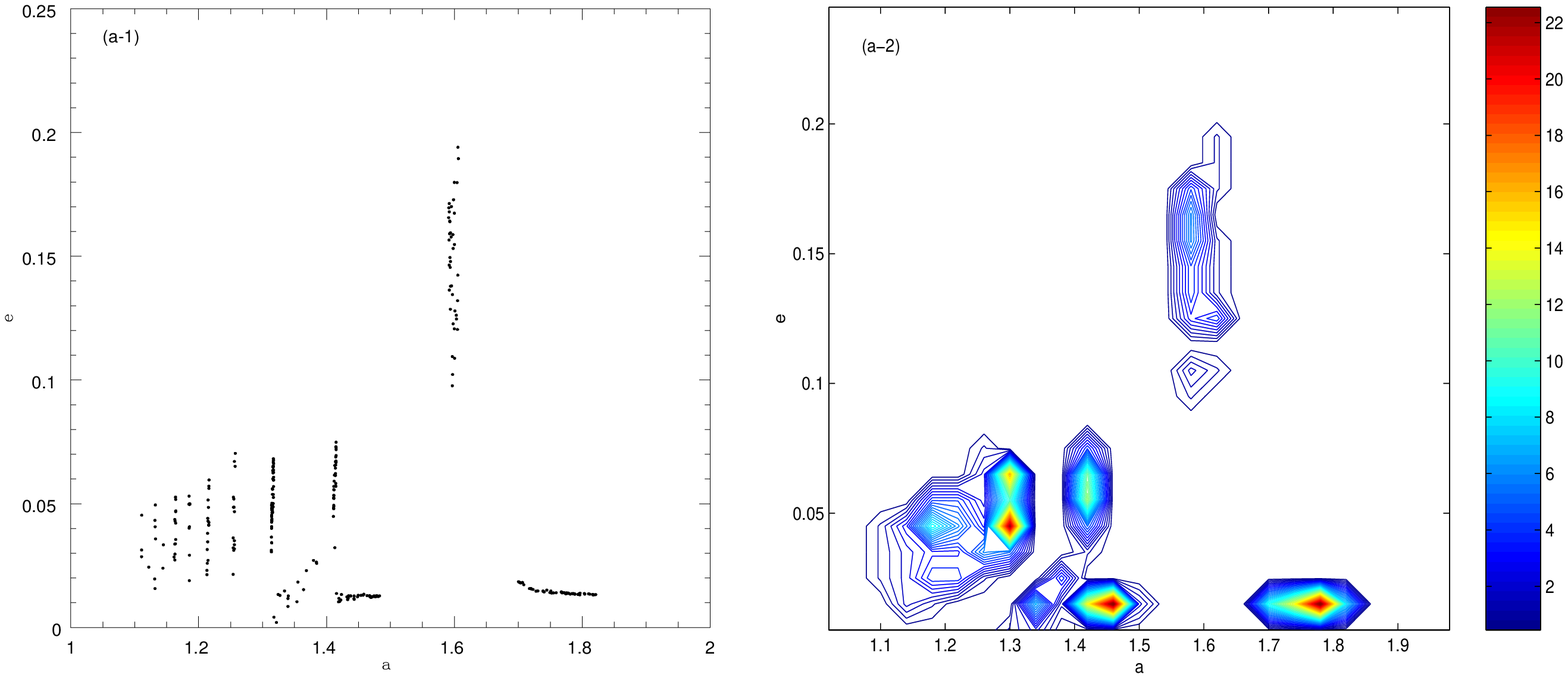}
\caption{The model A: the particle distribution in 
the $a$-$e$ plane at the end of simulation, 
i.e. $t=123200\pi$. The right color panel shows the number 
of particles at the particular area.}
\end{figure}
\clearpage
\begin{figure}[tbhp]
\epsfysize 6.5 in \epsffile{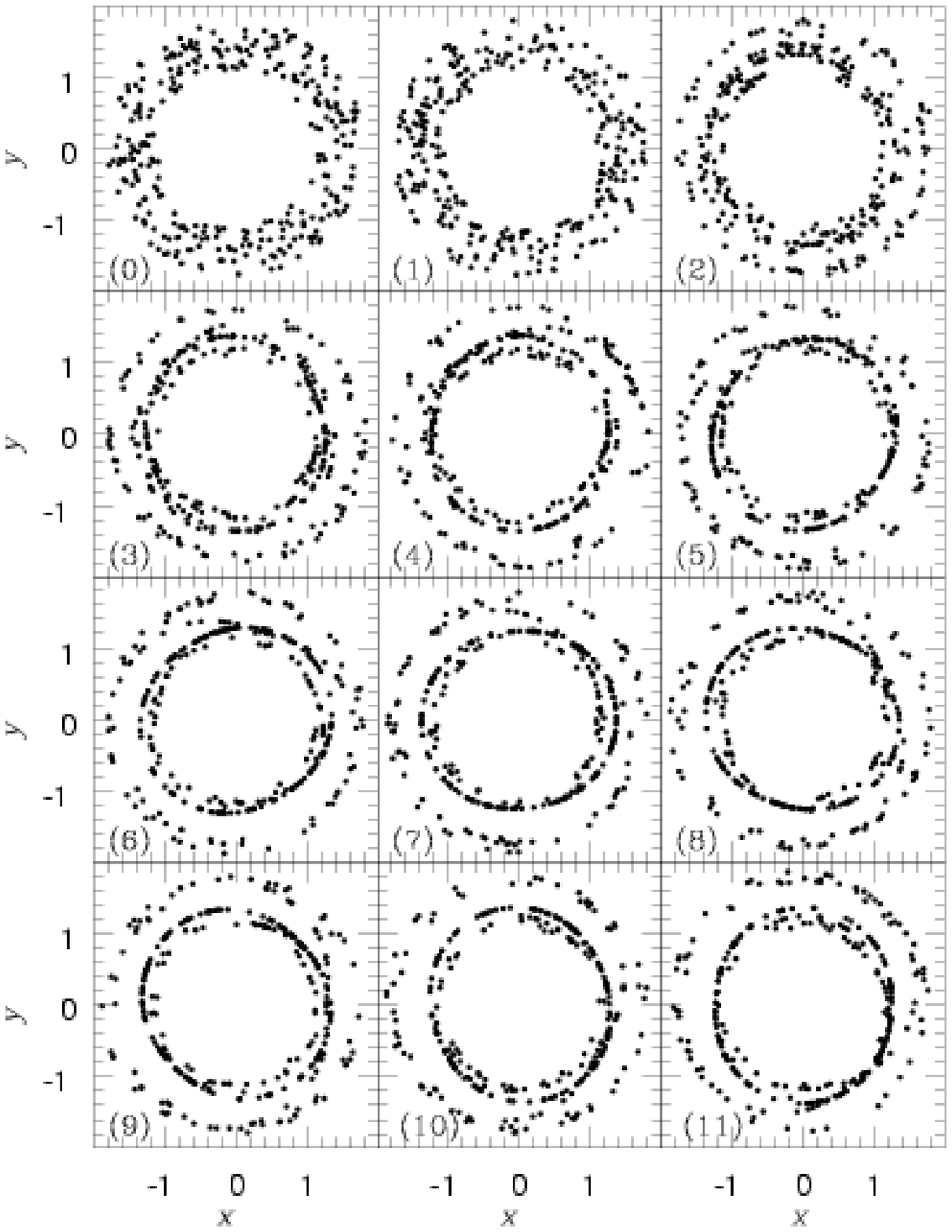}
\caption{The model B: the evolution of particle distributions in the 
$x-y$ plane.
The time between successive panels corresponds to $11200\pi$.}
\end{figure}
\clearpage
\begin{figure}[tbhp]
\epsfysize 6.5 in \epsffile{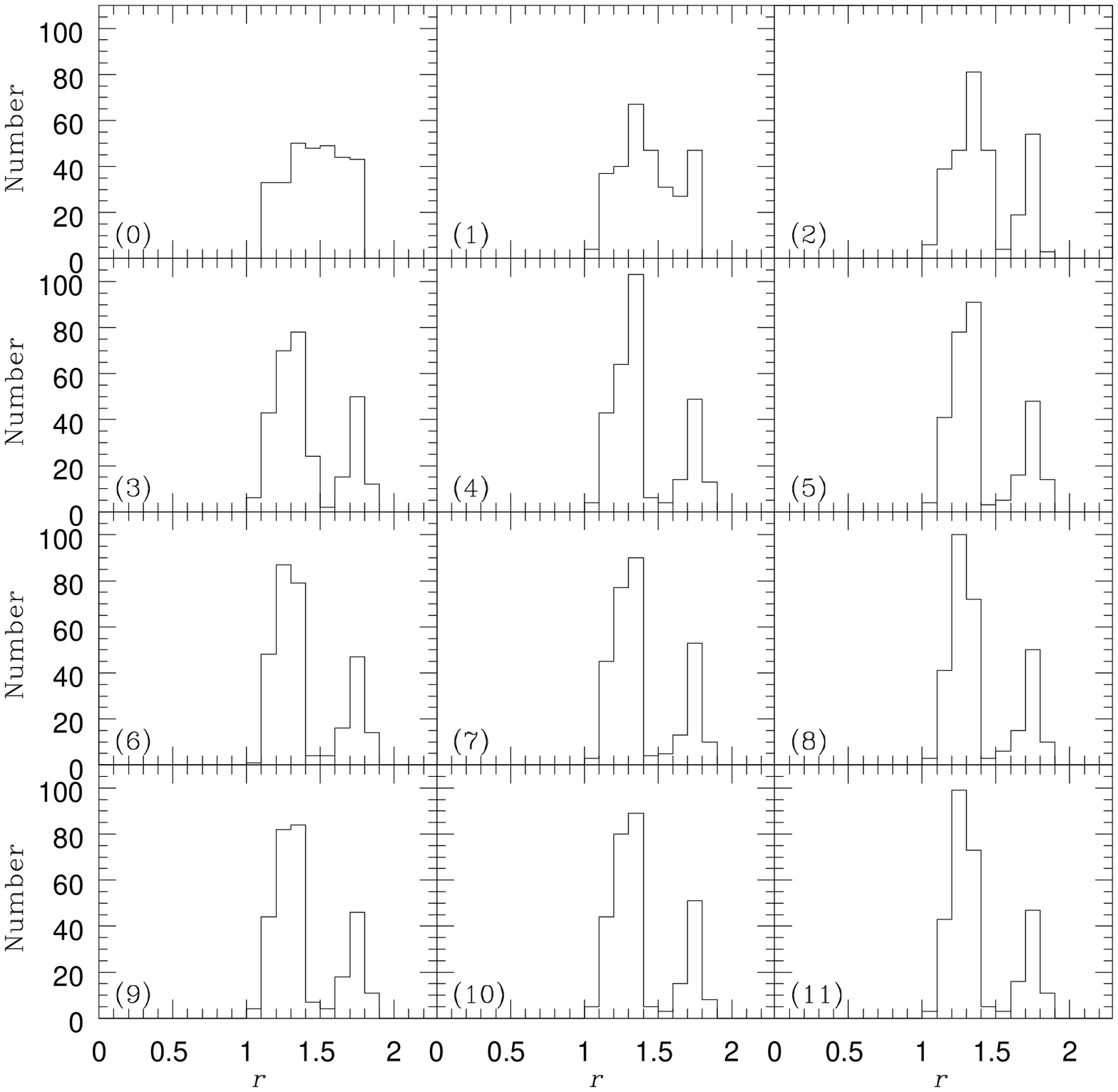}
\caption{The model B: the histograms of particle distributions 
in the radial coordinate.
The time between successive panels corresponds to $11200\pi$.}
\end{figure}
\clearpage
\begin{figure}[tbhp]
\epsfysize 3.0 in \epsffile{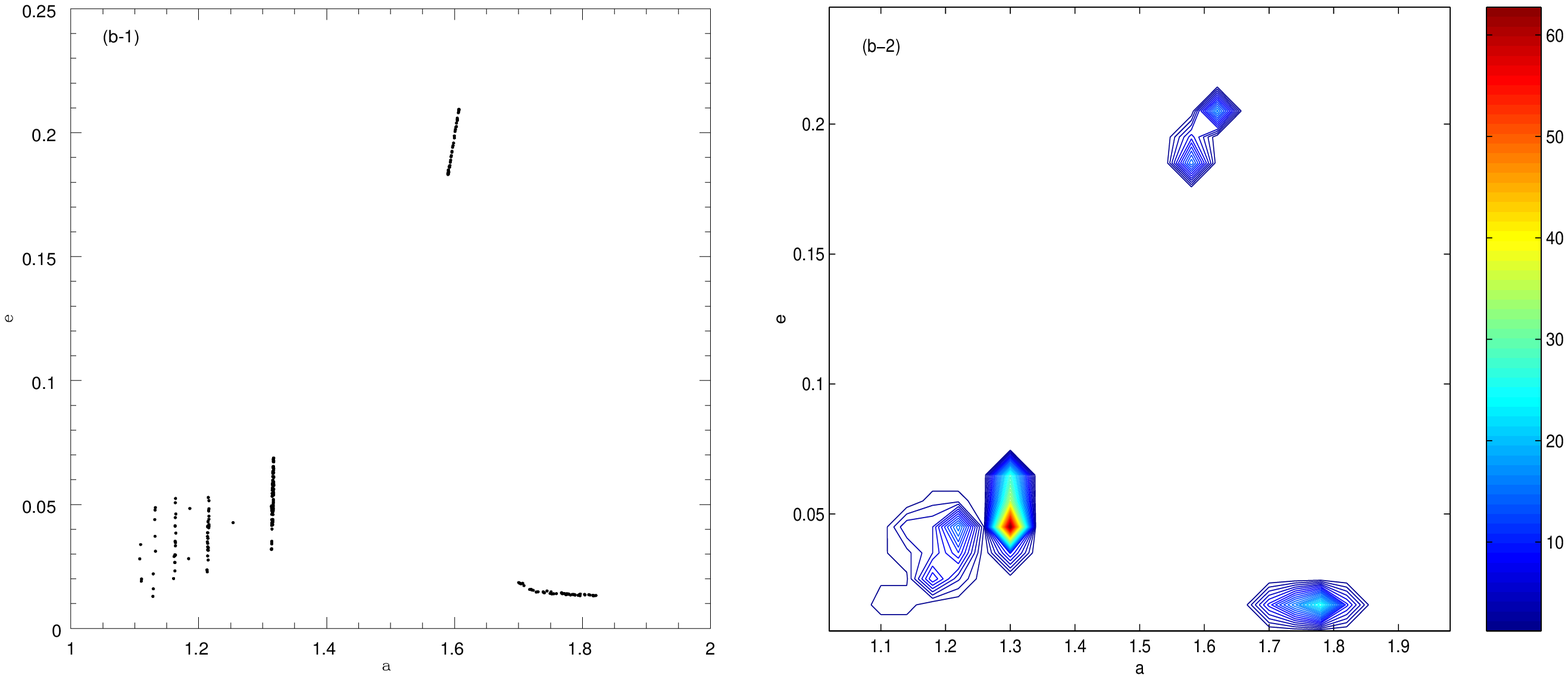}
\caption{The model B: the particle distribution in 
the $a$-$e$ plane at the end of simulation, 
i.e. $t=123200\pi$. The right color panel shows the number 
of particles at the particular area.}
\end{figure}
\clearpage
\begin{figure}[tbhp]
\epsfysize 6.5 in \epsffile{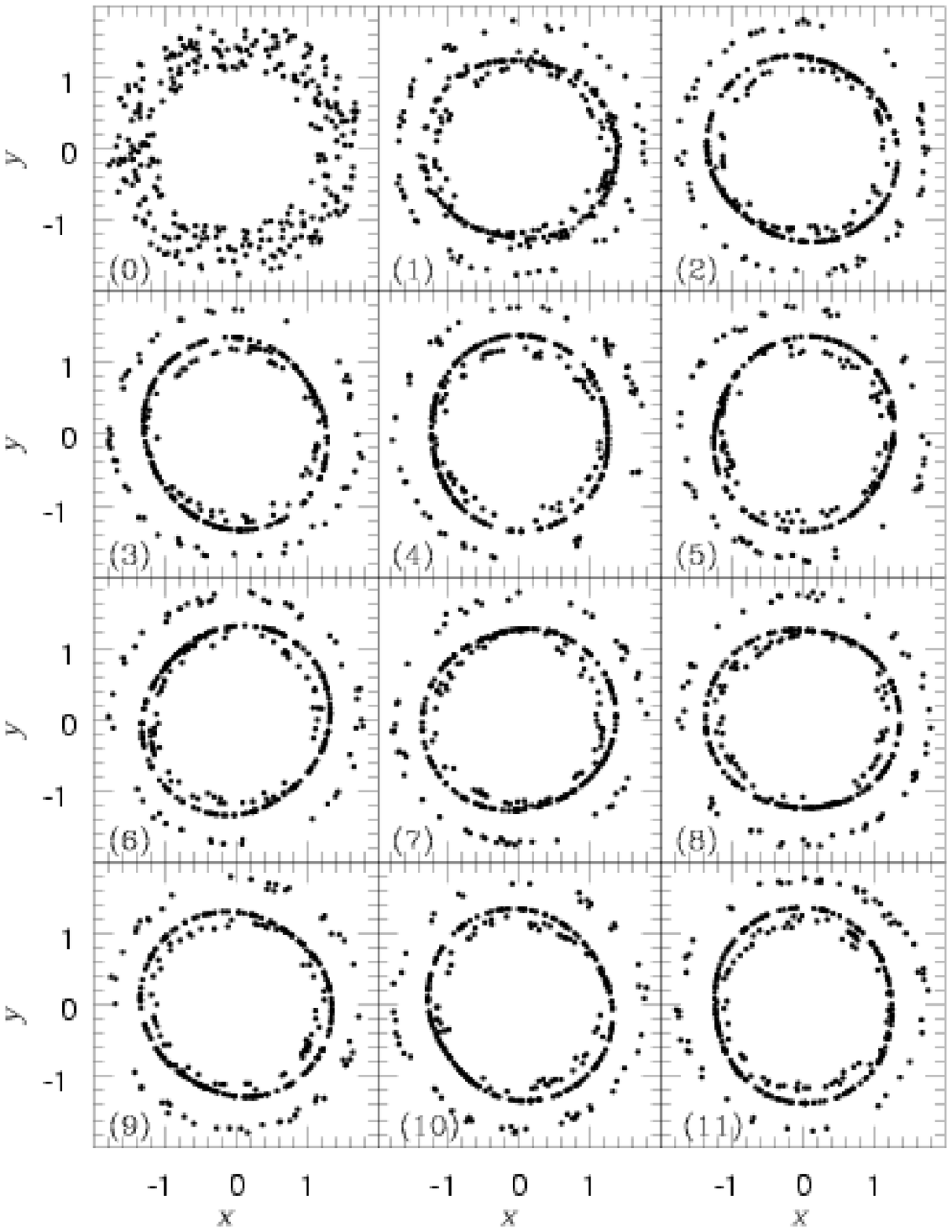}
\caption{The model C: the evolution of particle distributions in the 
$x-y$ plane.
The time between successive panels corresponds to $11200\pi$.}
\end{figure}
\clearpage
\begin{figure}[tbhp]
\epsfysize 6.5 in \epsffile{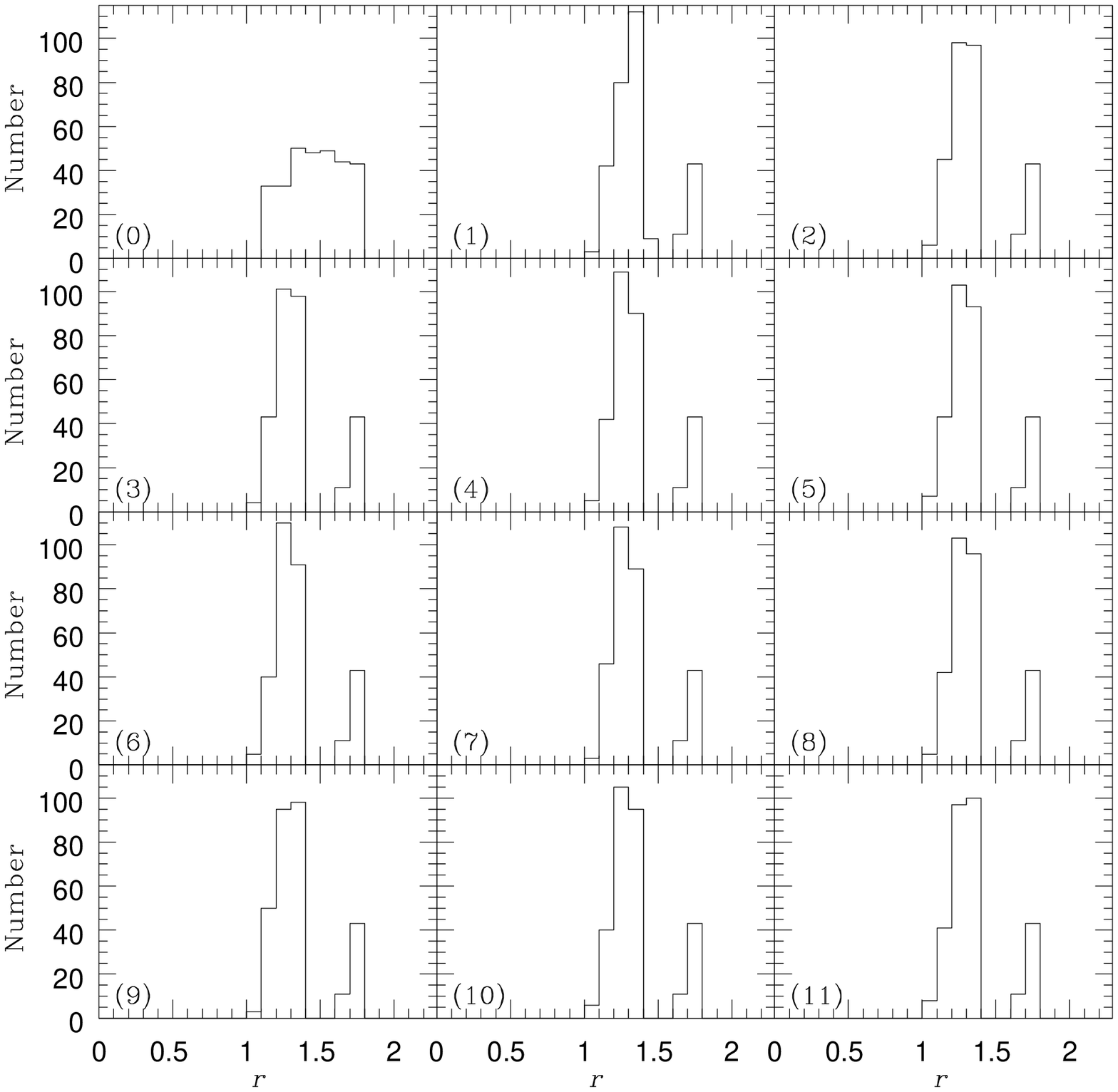}
\caption{The model C: the histograms of particle distributions 
in the radial coordinate.
The time between successive panels corresponds to $11200\pi$.}
\end{figure}
\clearpage
\begin{figure}[tbhp]
\epsfysize 3.0 in \epsffile{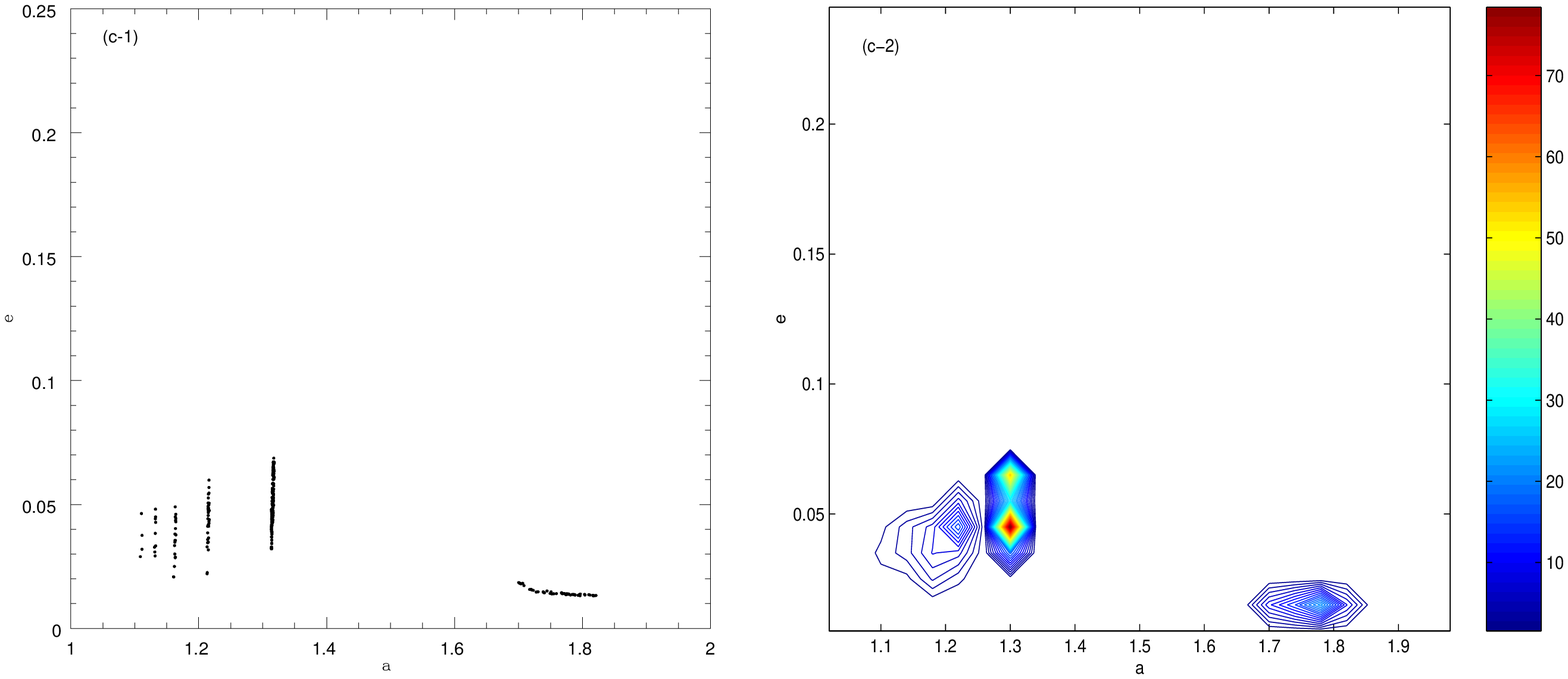}
\caption{The model C: the particle distribution in 
the $a$-$e$ plane at the end of simulation, 
i.e. $t=123200\pi$. The right color panel shows the number 
of particles at the particular area.}
\end{figure}

\clearpage
\begin{figure}[tbhp]
\epsfysize 6.5 in \epsffile{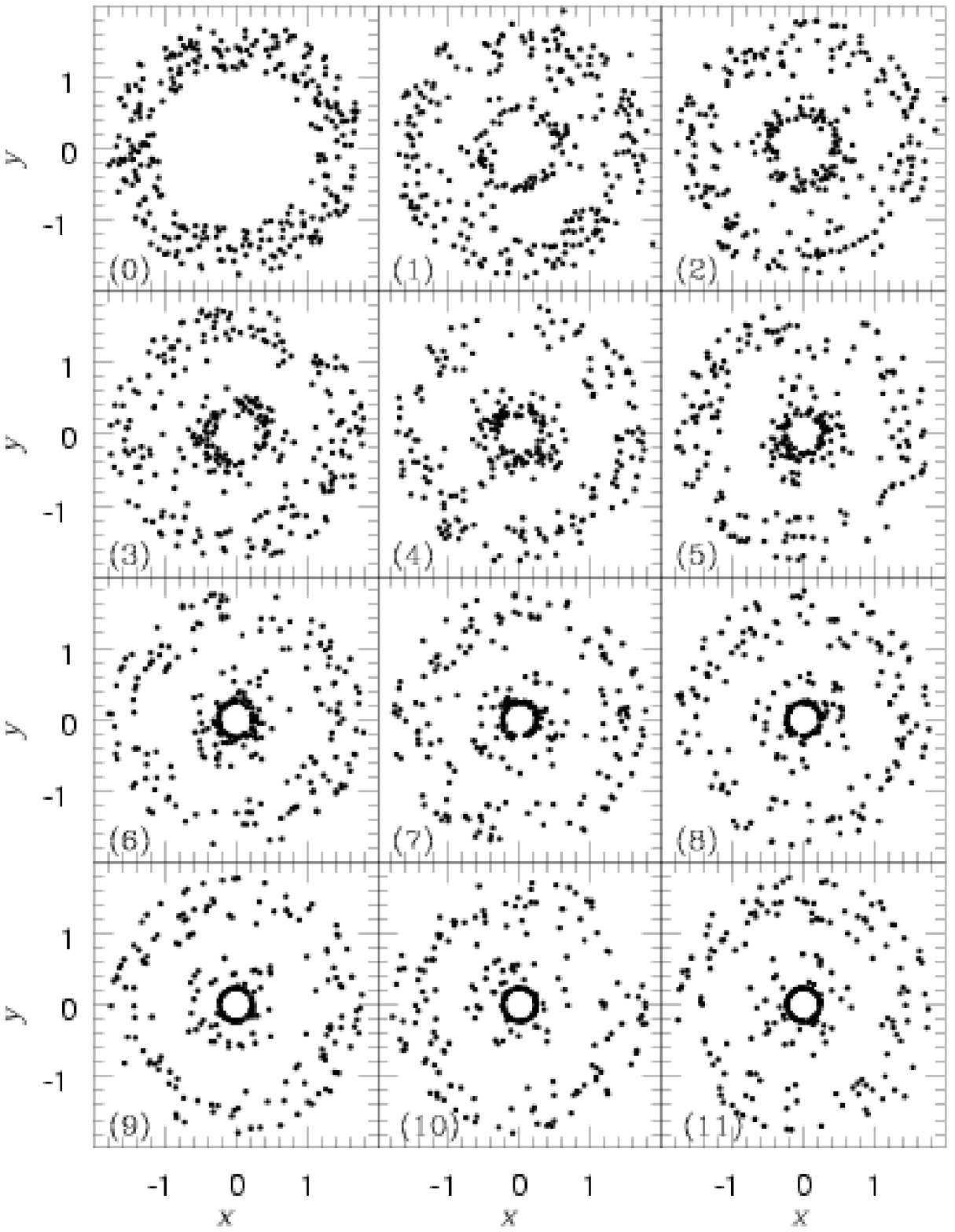}
\caption{The model D: the evolution of particle distributions in the 
$x-y$ plane.
The time between successive panels corresponds to $11200\pi$.}
\end{figure}
\clearpage
\begin{figure}[tbhp]
\epsfysize 6.5 in \epsffile{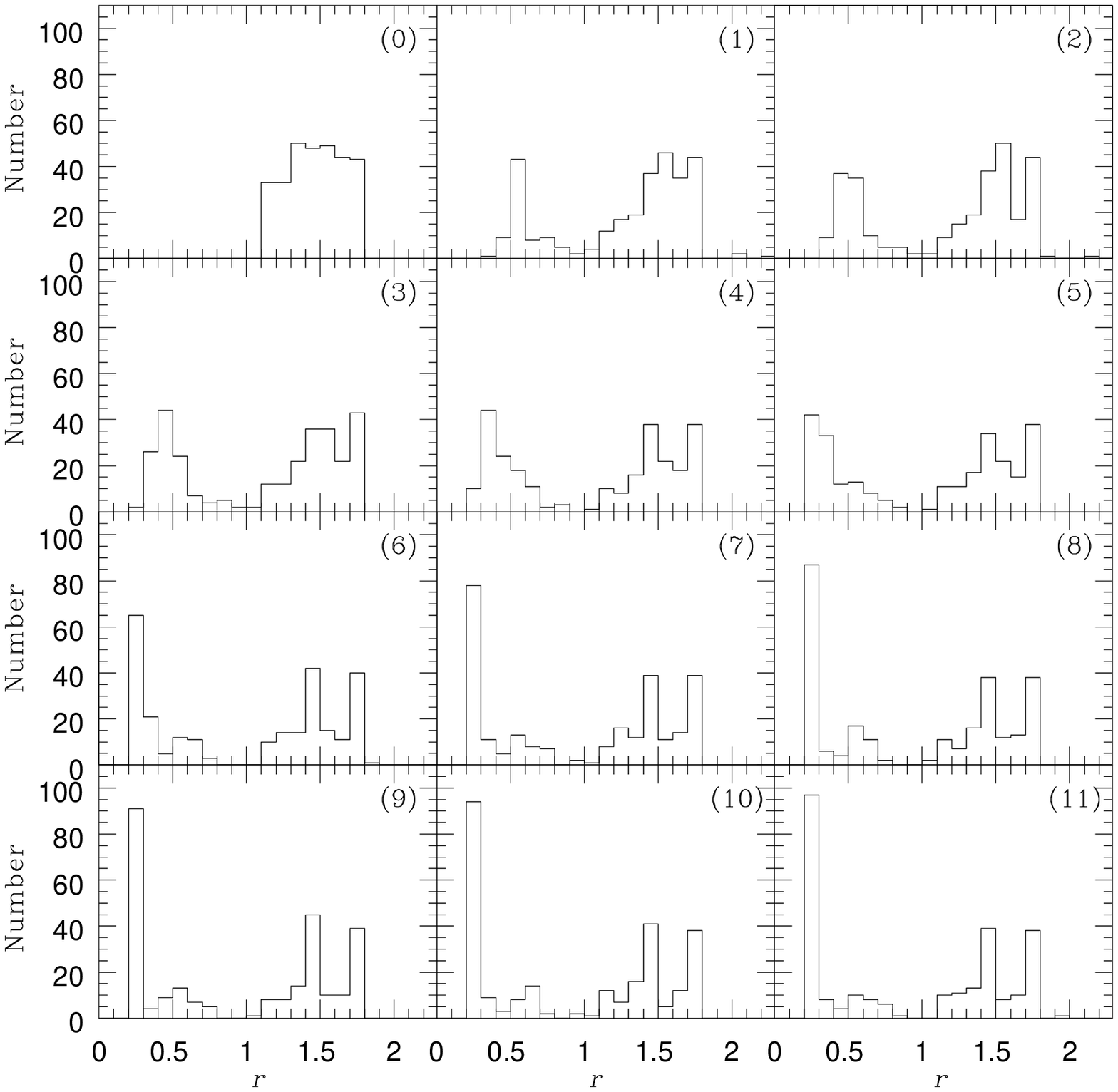}
\caption{The model D: the histograms of particle distributions 
in the radial coordinate.
The time between successive panels corresponds to $11200\pi$.}
\end{figure}
\clearpage
\begin{figure}[tbhp]
\epsfysize 6.5 in \epsffile{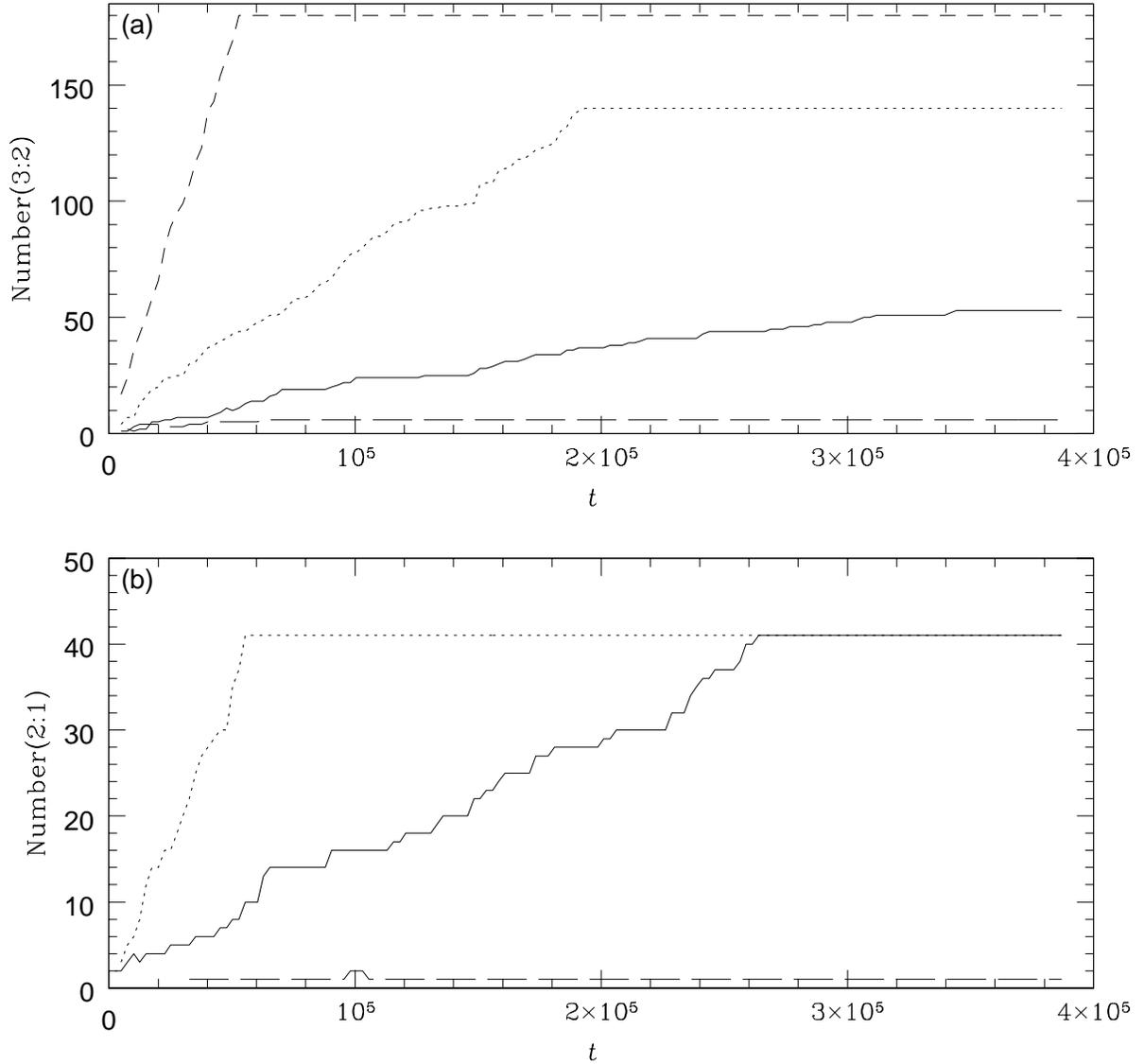}
\caption{The number of resonant particles as a function of time. 
The panel (a) is for the 3:2 resonance and panel (b)
is for the 2:1 resonance. The solid curves are for model A; the dotted curves 
are for model B; the dashed curve is for model C; the long dashed 
curves are for model D. Please note that there are only 3 curves in panel (b)
because no particle in the 2:1 resonance in model C.}
\end{figure}

\end{document}